\documentclass[final]{IEEEtran}
\pdfoutput=1
\usepackage{amsthm,amssymb,graphicx,multirow,amsmath,color,amsfonts}%,ulem}
\usepackage[caption=false,font=footnotesize]{subfig}
\usepackage[update,prepend]{epstopdf}
\usepackage[noadjust]{cite}
\usepackage[latin1]{inputenc}
\usepackage{tikz}
\usepackage{bbm} % for \mathbbm{1}
\usepackage{pdfpages}
\usepackage{multirow}
\usepackage{comment}
\usepackage{algorithm}
\usepackage{algorithmic}
\usepackage{url}
\def\BSTATE{\STATE\hskip-\ALG@thistlm}
\makeatother

\def\nb0{{\mathbf{0}}}
\def\nb1{{\mathbf{1}}}

\def\nbbA{{\mathbb{A}}}

\def\nbbS{{\mathbb{S}}}

\newtheorem{lemma}{Lemma}

\newtheorem{theorem}{Theorem}

\newtheorem{remark}{Remark}

\def\P{\mathbb{P}}

\def\g{\left.\right|}

\allowdisplaybreaks % Allows breaking of eqnarray over multiple pages (avoids unnecessary blanks in the document before eqnarray)

\usepackage{setspace}	% Remove in double column version. Also search for \setstretch in the body of the paper and comment these commands for double column

\begin{document}
%\pagenumbering{gobble}
\graphicspath{{./Figures/}}
\title{
AoI-optimal Joint Sampling and Updating for Wireless Powered Communication Systems%A Generalized Framework for Spatially Clustered RF-powered IoT Network
%Modeling and Analysis of RF-powered IoT Network using Poisson Cluster Process
%Performance Analysis of IoT RF-Powered by Secrecy Wireless Networks 
}
\author{
Mohamed A. Abd-Elmagid, Harpreet S. Dhillon, and Nikolaos Pappas
\thanks{M. A. Abd-Elmagid and H. S. Dhillon are with Wireless@VT, Department of ECE, Virginia Tech, Blacksburg, VA (Email: \{maelaziz,\ hdhillon\}@vt.edu). N. Pappas is with the Department of Science and Technology, Link\"{o}ping University, SE-60174 Norrk\"{o}ping, Sweden (Email: nikolaos.pappas@liu.se). The support of the U.S. NSF (Grant CPS-1739642) is gratefully acknowledged.}
\vspace{-5mm}
}

%\author{
%Author 1, Author 2, and Author 3
%\vspace{-5mm}
%}

\maketitle

\begin{abstract}
  This paper characterizes the structure of the Age of Information \textit{(AoI)-optimal policy} in wireless powered communication systems while accounting for the \textit{time and energy costs} of generating status updates at the source nodes. In particular, for a single source-destination pair in which a radio frequency (RF)-powered source sends status updates about some physical process to a destination node, we minimize the long-term average AoI at the destination node. The problem is modeled as an average cost \textit{Markov Decision Process} (MDP) in which, the generation times of status updates at the source, the transmissions of status updates from the source to the destination, and the wireless energy transfer (WET) are jointly optimized. After proving the monotonicity property of the value function associated with the MDP, we analytically demonstrate that the AoI-optimal policy has a \textit{threshold-based} structure w.r.t. the state variables. Our numerical results verify the analytical findings and reveal the impact of state variables on the structure of the AoI-optimal policy. Our results also demonstrate the impact of system design parameters on the optimal achievable average AoI as well as the superiority of our proposed \textit{joint sampling and updating policy} w.r.t. the \textit{generate-at-will} policy.
\end{abstract}
%\begin{IEEEkeywords}
%Age of Information, Internet of Things, Markov Decision Process, Wireless Energy Transfer.
%\end{IEEEkeywords}
%\vspace{-0.5cm}
\section{Introduction}\label{sec:intro}
AoI provides a rigorous way of quantifying the freshness of information about a physical process at a destination node based on the status updates it receives from a source node \cite{abd2018role}. In \cite{kaul2012real}, AoI was first defined as the time elapsed between the generation of a status update at the source and its reception at the destination. Since then, AoI has been extensively used to quantify the performance of various communication networks that deal with time-sensitive information, including multi-hop networks \cite{talak2017}, multicast networks \cite{Buyukates_ulu}, broadcast networks \cite{kadota2016,bastopcu2020should}, and ultra-reliable low-latency vehicular networks \cite{abdel2018ultra}. Interested readers are advised to refer to \cite{kosta2017age_mono,sun2019age} for comprehensive surveys.

Recently, the concept of AoI has been argued to have an important role in designing {\it freshness-aware Internet of Things} (IoT) networks (which can enable a broad range of real-time applications) \cite{gu2019timely,abd2018average,zhou2018joint,AbdElmagid2019Globecom_b,Praful_GC1}. A common assumption in most of the literature on AoI is to neglect the costs of generating status updates, however, the IoT devices (source nodes in the context of AoI setting) are currently expected to perform {\it sophisticated tasks} while generating status updates \cite{zhou2018joint,fountoulakis2020optimal}. In that sense, it is crucial to incorporate the energy and time costs of generating status updates in the design of future freshness-aware IoT networks. To further enable a sustainable operation of such networks, RF energy harvesting has emerged as a promising solution for charging low-power IoT devices \cite{abd2018coverage}. 
 %due to its cost-efficient implementation
 In particular, the ubiquity of RF signals even at
hard-to-reach places makes them more suitable to power IoT devices than other popular sources of energy harvesting, such as solar or wind. In addition, the implementation of RF energy harvesting modules is usually cost efficient, which is another important aspect of the deployment of IoT devices.
 %In particular, the ubiquity of Due to its ubiquity and cost efficient implementation, radio frequecy (RF) energyharvesting has quickly emerged as an appealing solution for powering IoT devices (majority ofwhich are tiny devices, such as sensors, with very low energy requirement)
The main focus of this paper is to investigate the structural properties of the AoI-optimal joint sampling and updating policy for freshness-aware RF-powered IoT networks. 
 
The AoI-optimal policy for an energy harvesting source has already been investigated under various system settings \cite{yates2015lazy,10,8,2_2,2_3,ceran2019reinforcement,George2019,ozel2020timely}. The energy harvesting process is commonly modeled as an independent external stochastic process. However, when the source is assumed to be RF-powered, the harvested energy depends on the channel state information (CSI) and its variation over time, which makes the characterization of the AoI-optimal policies very challenging. It is worth noting that \cite{2_5,2_8,abd2019tcom} have very recently explored the AoI-optimal policy in wireless powered communication systems. However, none of the proposed policies took into account the time and energy costs of generating status updates at the source. In addition, \cite{2_5,2_8} did not incorporate the evolution of the battery level at the source and the variation of CSI over time in the process of decision-making.
This paper makes the first attempt to analytically characterize the structural properties of the AoI-optimal joint sampling and updating policy while: {\it i) considering the dynamics of battery level, AoI, and CSI, and ii) accounting for the costs of generating status updates in the process of decision-making.} 
%This, in turn, means that the proposed AoI-optimal policies in \cite{yates2015lazy,2_3,10} are no longer applicable to such system settings since one needs to explicitly incorporate the statistics of CSI in the process of decision-making.
%To quantify freshness of information at the destination node, we use AoI as a performance metric. The authors of \cite{kaul2012real} introduced the concept of AoI and characterized average AoI for a simple queueing-theoretic model. Building on this,  a series of works \cite{yates2012real,costa2016age,Modiano2015,kosta2017age} focused on characterizing the average AoI and its variations (e.g., Peak Age-of-Information \cite{costa2016age} and Value of Information of Update \cite{kosta2017age}) for adaptations of the queueing model studied in \cite{kaul2012real}. Another direction of research \cite{sun2017update,ABedewy2016,abd2018average,113882,talak2018optimizing,AbdElmagid2019Globecom_b,yates2015lazy,2_3,10,2_2,3,8,2_5,2_8} focused on applying various tools from optimization theory to characterize AoI-optimal transmission policies for different communication systems that deal with time critical information.

{\it Contributions.} Our main contribution is the analytical characterization of the structure of the AoI-optimal policy for an RF-powered single source-destination pair system setup while incorporating the time and energy costs for generating status updates at the source. In particular, we model the problem as an average cost MDP\footnote{The theory of MDPs is useful for problems in which the objective is to obtain an optimal mapping between the system state and action spaces. It also allows one to account for the temporal variations of the system state variables in the process of decision-making.} with finite state and action spaces for which its corresponding value function is shown to be a monotonic function w.r.t. the state variables. Using this property, the AoI-optimal policy is proven to have a threshold-based structure w.r.t. different state variables. Our numerical results verify our analytical findings and reveal the impact of state variables as well as the energy required for generating a status update at the source on the structure of the AoI-optimal policy. Our results also demonstrate that the optimal achievable average AoI by our proposed joint sampling and updating policy significantly outperforms the achievable average AoI by the generate-at-will policy.
\begin{figure}[t!]
\centering
\includegraphics[width=0.76\columnwidth]{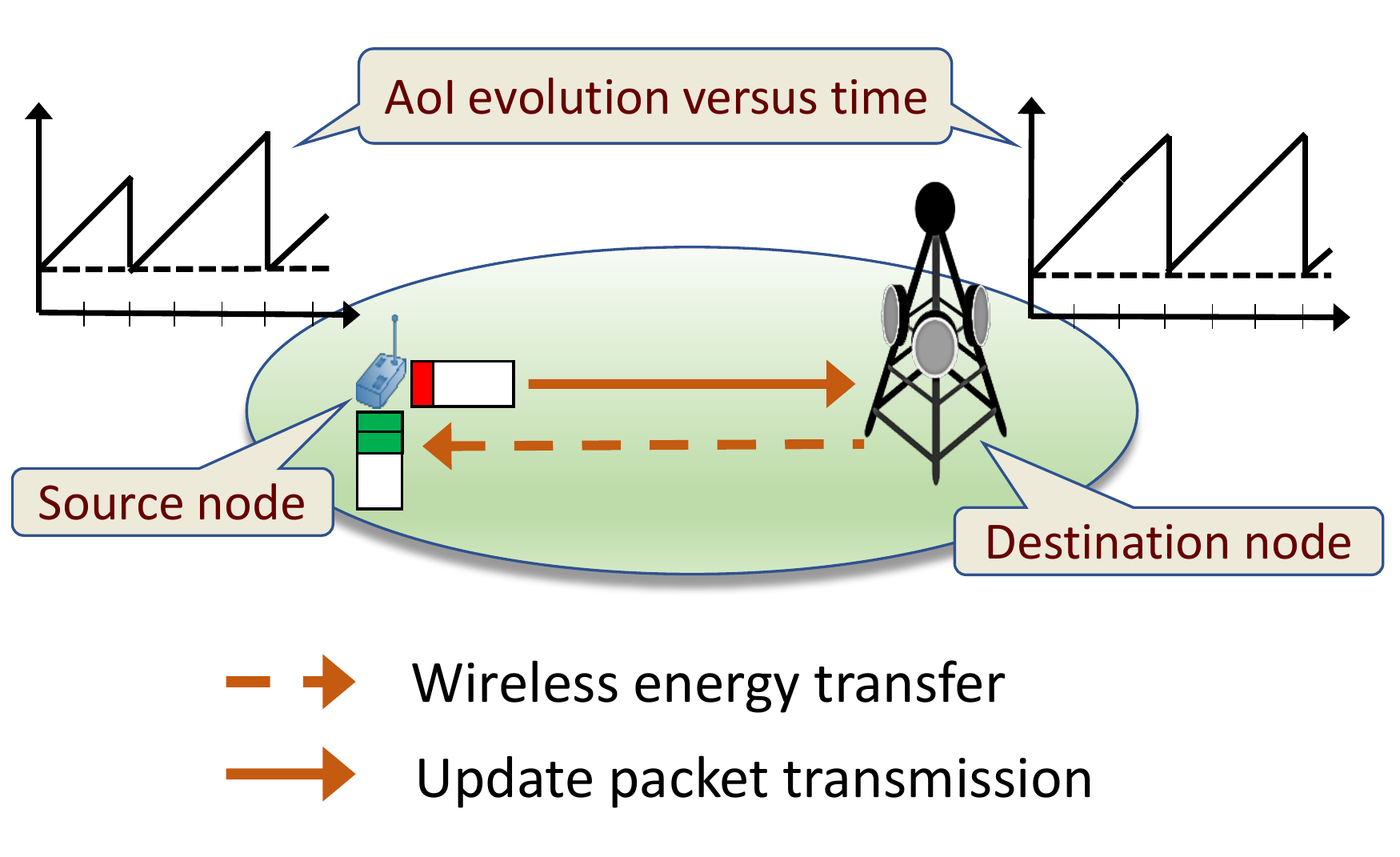}
\caption{An illustration of the system setup.}
\label{sys_model}
\end{figure}
\section{System Model and Problem Formulation}\label{sec:Model}
\subsection{Network Model}
We consider a single source-destination pair model in which the source contains: i) a sensor that keeps sampling the real-time status of a physical process and ii) a transmitter that sends status update packets about the observed process to the destination, as shown in Fig. \ref{sys_model}. 
%In an IoT setup, an aggregator (source) collects status updates from nearby devices and transmits to a cellular base station (destination). 
Since the single source-destination pair model may actually be sufficient to study a diverse set of applications \cite{kaul2012real} (e.g., safety of an intelligent transportation system, predicting and controlling forest fires, and efficient energy utilization in future smart homes), our analysis in this paper will be of interest in many applications. The scenario of having multiple source nodes is left as a promising direction of future work. 

We assume that the source node may perform sophisticated sampling tasks, e.g., initial feature extraction and pre-classification using machine learning tools \cite{zhou2018joint}. Hence, unlike most of the existing literature, the time and energy costs of generating an update packet at the source node cannot be neglected. While the destination node is assumed to be always connected to the power grid, the source node is powered through WET by the destination node. 
%Particularly, the source harvests energy from the RF signals transmitted by the destination in the downlink, and stores it in a battery with finite capacity $B_{\rm max}$~joules.
Particularly, the destination node transmits RF signals in the downlink to charge the source node. The energy harvested by the source node is then stored in its battery, which has a finite capacity of $B_{\rm max}$~joules. The source and destination nodes share the same channel and they have a single antenna each.
%The source and destination nodes are assumed to operate over the same frequency channel, and have a single antenna each.
Hence, the source can either harvest energy or transmit data at a given time instant.

%We assume discrete time and the time slots are of equal size.
We assume discrete time and the time slots are of equal size. Let $B(n)$, $A(n)$ and $\tau(n)$ denote the amount of available energy in the battery at the source node, the AoI at the destination node, and the time passed since the generation instant of the current update packet available at the source node (i.e., the AoI of the status updates at the source node), respectively, at the beginning of time slot $n$. Denote by $h(n)$ and $g(n)$ the uplink and downlink channel power gains between the source and destination nodes over slot $n$, respectively. We assume that the channels are influenced by quasi-static flat fading. This, in turn, means that the channels are fixed over a time slot, and independently vary from one slot to another.
%The channels are assumed to be affected by quasi-static flat fading, i.e., they remain constant over a time slot but change independently from one slot to another. 
%The destination node has a perfect knowledge about the channel power gains in the current time slot, and only a statistical knowledge for future slots.
%%
%The location of the source node is known {\em a priori}, and hence its average channel power gains are pre-estimated and known at the destination node. Specifically, the destination node has perfect knowledge about the channel power gains in the current time slot, and only a statistical knowledge for future slots. This is a reasonable assumption for many IoT applications.
%%
%We assume that $A(n)$ is upper bounded by a finite value $A_{\rm max}$ which can be chosen to be arbitrarily large, i.e., $A(n) \in \{1,2,\cdots,A_{\rm max}\}$. When AoI reaches $A_{\rm max}$, it means that the information is too stale to be of any use at the destination node. 
%%
% We employ the concept of AoI as a measure of freshness of information at the destination
\subsection{State and Action Spaces}
The state of the system at slot $n$ can be expressed as $s(n) \triangleq \left(B\left(n\right), A\left(n\right), \tau\left(n\right), h\left(n\right), g\left(n\right) \right) \in \nbbS$; where $\nbbS$ is the state space which contains all the combinations of the system state variables. We also assume that the state variables can take discrete values\footnote{Note that constructing a finite state space of an MDP by discretizing the state variables and/or defining upper bounds on their maximum values is very common in the literature to obtain the optimal policy numerically as well as characterize its structure properties analytically using standard optimization techniques such as the Value Iteration Algorithm (VIA) or Policy Iteration Algorithm (PIA). See~\cite{zhou2018joint,ceran2019reinforcement,George2019,AlessWPCNs} for representative examples.} and obtain a lower bound to the performance of the continuous system (as it will be clear in the sequel). In particular, we have $B(n) \in \{0,1,\cdots,b_{\rm max}\}$ where $b_{\rm max}$ denotes the battery capacity, such that each energy quantum in the battery is equivalent to $e_{\rm q} = \frac{B_{\rm max}}{b_{\rm max}}$ joules.
%denotes the size of the battery, such that each slot in the battery is equivalent to $e_{\rm q} = \frac{B_{\rm max}}{b_{\rm max}}$ joules.
Note that both the energy consumed from the battery for an update packet transmission and the harvested energy need to be expressed in terms of the energy quanta. In addition, if the channel power gains are originally modeled using continuous random variables, we discretize them into a finite number of intervals whose probabilities are determined from the probability density function (PDF) of the fading gain. In particular, each interval is then represented by a discrete level of channel power gain which has the same probability as that of this interval. Without loss of generality, we also assume that $A(n)$ $(\tau(n))$ is upper bounded by a finite value $A_{\rm max}$ $(\tau_{\rm max})$ which can be chosen to be arbitrarily large \cite{zhou2018joint,ceran2019reinforcement}.
%, i.e., $A(n) \in \{1,2,\cdots,A_{\rm max}\}$ and $\tau(n) \in \{1,2,\cdots,\tau_{\rm max}\}$. 
When $A(n)$ reaches $A_{\rm max}$, it means that the available information at the destination node is too stale to be of any use. 
%Note that this also does not restrict the generality of our analysis/problem formulation since these upper bound values can be chosen to be arbitrarily large.

Based on $s(n)$, two actions are decided at slot $n$: i) the first action $a_1(n) \in \nbbA_1 \triangleq \{S, I\}$ determines whether the source generates a new update packet in slot $n$ or not, and ii) the second action $a_2(n) \in \nbbA_2 \triangleq \{T, H\}$ determines whether slot $n$ is allocated for an update packet transmission from the source to the destination or WET by the destination. Specifically, when $a_1(n) = S$, a new update packet is generated by the source, which replaces the currently available one, if any, since there is no benefit of sending out-of-date packets to the destination. We also consider that generating an update packet takes one time slot (as a {\it time cost}) and requires an amount of energy $E^{\rm S}$ (as {\it energy cost} expressed in energy quanta). When $a_2(n) = T$, the source sends its currently available packet (that was generated from $\tau(n)$ time slots) to the destination. The required energy for a packet transmission of size $M$ bits in slot $n$, according to Shannon's formula, is $E^{\rm T}(n) = \frac{\sigma^2}{h(n)}\left(2^{M/W} - 1\right)$, where $\sigma^2$ is the noise power at the destination and $W$ is the channel bandwidth. When $a_{2}(n) = H$, slot $n$ is allocated for WET by the destination to charge the battery at the source. We consider a practical non-linear energy harvesting model \cite{boshkovska2015practical} such that the energy harvested by the source is given by 
\begin{align}\label{eq:harv_energy}
E^{\rm H}(n) = \frac{P_{\rm max}\left(1 - {\rm exp}\left[- a P_{\rm rec}\left(n\right)\right]\right)}{1 + {\rm exp}\left[- a \left(P_{\rm rec}\left(n\right) - b\right)\right]}, 
\end{align}
where $a$ and $b$ are constants representing the steepness and the inflexion point of the curve that describes the input-output power conversion, $P_{\rm max}$ is the maximum power that can be harvested through a particular circuit configuration, and $P_{\rm rec}(n) = P_{\rm t} g(n)$ such that $P_{\rm t}$ is the average transmit power by the destination. Hence the system action at slot $n$ can be expressed as $a\left(n\right) = \left(a_1\left(n\right), a_2\left(n\right)\right) \in \nbbA = \nbbA_1 \times \nbbA_2$, where $\nbbA$ is the action space of the system. 

Note that the system state is assumed to be available at the destination node at the beginning of each time slot to take decisions. In particular, we assume that the location of the source node is known {\em a priori}, and hence the average channel power gains are pre-estimated and known at the destination node. In particular, at the beginning of an arbitrary time slot, the destination node has perfect knowledge about the channel power gains in that slot, and only a statistical knowledge for future slots \cite{AlessWPCNs}. Further, given some initial values for the remaining system state parameters (i.e., $B(0)$, $\tau(0)$ and $A(0)$), the destination node updates their values based on the action taken at each time slot. More specifically, $B(n + 1)$ can be expressed as a function of the system action at slot $n$ $(a(n))$ as
\begin{align}\label{eq:batt_evol}
\begin{cases}
\begin{aligned}
&B(n) - \left \lceil E^{\rm T}(n)/ e_{\rm q} \right \rceil, && \text{if}\; a(n) = (I,T),\\
&B(n) - E^{\rm S} - \left \lceil E^{\rm T}(n)/ e_{\rm q} \right \rceil, && \text{if}\;a(n) = (S,T),\\
&{\rm min} \left \{b_{\rm max}, B(n)+ \left \lfloor E^{\rm H}(n)/ e_{\rm q} \right \rfloor \right \}, && \text{if}\;a(n) = (I,H),\\
&{\rm min} \left \{b_{\rm max}, B(n) - E^{\rm S} + \left \lfloor E^{\rm H}(n)/ e_{\rm q} \right \rfloor \right \}, && \text{if}\;a(n) = (S,H),
\end{aligned}
\end{cases}
\end{align}
where we used the ceiling and floor with $E^{\rm T}(n)$ and $E^{\rm H}(n)$, respectively. Thus, we obtain a lower bound to the performance of the original continuous system. An upper bound can be obtained by reversing the ceiling and floor operators. Let $\nbbA\left(s\left(n\right)\right)$ denote the action space associated with state $s(n)$, i.e., $\nbbA\left(s\left(n\right)\right)$ contains the possible actions that can be taken at $s(n)$. We assume that $a(n) \in \nbbA\left(s\left(n\right)\right)$ only if $B(n)$ is greater than the energy required for taking action $a(n)$, hence we always have $B(n + 1) \geq 0$. Furthermore, $A(n+1)$ and $\tau(n+1)$ can be expressed, respectively, as
\begin{align}\label{eq:AoI_evol}
A(n + 1) = \begin{cases}
\begin{aligned}
&{\rm min}\left \{A_{\rm max}, \tau(n)+1 \right \},\; &&\text{if}\; a(n) = \left(a_1\left(n\right),T\right)\\
&{\rm min}\left \{A_{\rm max}, A(n)+1 \right \},\; &&\text{otherwise}.
\end{aligned}
\end{cases}
\end{align}
\begin{align}\label{eq:tau_evol}
\tau(n + 1) = \begin{cases}
\begin{aligned}
&1,\; &&\text{if}\; a(n) = \left(S,a_2\left(n\right)\right),\\
&{\rm min}\left \{\tau_{\rm max}, \tau(n)+1 \right \},\; &&\text{otherwise},
\end{aligned}
\end{cases}
\end{align}
where $a(n) = (a_1(n),T)$ means that $a(n) \in \{(I,T),(S,T)\}$. This also applies to $(S,a_2(n))$ w.r.t. $a_2(n)$.
%To help visualize (\ref{eq:AoI_evol}), Fig. \ref{f:AoI_illustration} shows the AoI evolution as a function of actions taken over time when $A_{\rm max} = 4$.
%\begin{figure}[t!]
%\centering
%\includegraphics[width=0.55\columnwidth]{AoI_illustration_1.pdf}
%\caption{AoI evolution vs. time when $A_{\rm max} = 4$.}
%\label{f:AoI_illustration}
%\end{figure} 
\subsection{Problem Formulation}
A policy is a mapping from the system state space to the system action space. Under a policy $\pi$, the long-term average AoI at the destination with initial state $s(0)$ is given by
\begin{align}\label{average_AoI}
\bar{A}^{\pi} \triangleq \limsup_{N\to \infty} \frac{1}{N+1} \sum_{n = 0}^N \mathbb{E}\left[A(n) \g s(0)\right].
\end{align}

We take the expectation w.r.t. the channel conditions and policy in (\ref{average_AoI}). We then aim at finding the policy $\pi^*$ that achieves the minimum average AoI, i.e., 
\begin{align}\label{Optim_prob}
\pi^\star = {\rm arg}\; \underset{\pi}{\rm min} \;\bar{A}^{\pi}.
\end{align}
%\normalsize

Owing to the independence of channel power gains over time and the nature of the dynamics of remaining state variables, as described by (\ref{eq:batt_evol})-(\ref{eq:tau_evol}), the problem can be modeled as an MDP. Recall that the system state space is finite (the state variables are discretized) and the system action space is clearly finite as well. In this case, the MDP at hand is a \textit{finite-state finite-action MDP}, for which there exists an optimal stationary deterministic policy (i.e., we take a deterministic action at each state that is fixed over time) that can be obtained using the VIA or PIA \cite{bertsekas2011dynamic}. Therefore, in the sequel, we omit the time index and explore this stationary deterministic policy. In the next section, we characterize the AoI-optimal policy $\pi^\star$ and derive its structural properties.

\section{Analysis of the AoI-optimal Policy}
\subsection{Optimal Policy Characterization}
Given a stationary deterministic policy $\pi$, the probability of moving from state $s = (B,A,\tau,h,g)$ to state $s'=(B',A',\tau',h',g')$ can be expressed as
\begin{align}\label{transprob}
&\nonumber \P\left(s' \g s, \pi(s)\right)  \triangleq \P\left(B', A', \tau', h', g' \g B, A, \tau, h, g, \pi(s)\right)\\
\nonumber & \overset{({\rm a})}{=} \P\left(B', A', \tau' \g B, A, \tau, h, g, \pi(s) \right) \P(h') \P(g')\\
& \overset{({\rm b})}{=} C \P\left(B' \g B, h, g, \pi(s)\right) \P\left(A' \g A,\tau, \pi(s) \right) \P\left(\tau' \g \tau, \pi(s) \right),
\end{align}
where $\pi(s)$ denotes the action taken at state $s$ according to $\pi$, $\P(h')$ and $\P(g')$ denote the probability mass functions for the uplink and downlink channel power gains, and $C = \P\left(h'\right) \P\left(g'\right)$. Step (a) follows since the channel power gains are independent over time from each other and from other random variables. Note that for the case of a Markovian fading channel model, the conditional probabilities $\P(h' \g h)$ and $\P(g' \g g)$ will replace $\P(h')$ and $\P(g')$, respectively. These conditional probabilities are determined according to the Markovian fading channel model considered in the problem. However, all our analytical results regarding the structure of the AoI-optimal policy (derived in the next subsection) will remain the same. Step (b) follows due to the fact that given $s$ and $\pi(s)$, we can obtain $B'$, $A'$ and $\tau'$ in a deterministic way separately from each other using (\ref{eq:batt_evol})-(\ref{eq:tau_evol}). 
%In particular, $B'$, $A'$ and $\tau'$ can be determined, respectively, as
%\begin{align}\label{eq:batt_transprob}
%\P\left(B' \g B, g, h, \pi(s)\right) =
%\begin{cases}
%\begin{aligned}
%&\mathbbm{1} \left( B' = B - \left \lceil E^{\rm T}/ e_{\rm q} \right \rceil \right), && \text{if}\; \pi(s) = (I,T),\\
%&\mathbbm{1} \left( B' = B - E^{\rm S} - \left \lceil E^{\rm T}/ e_{\rm q} \right \rceil \right), && \text{if}\;\pi(s) = (S,T),\\
%&\mathbbm{1} \left( B' = {\rm min} \left \{b_{\rm max}, B+ \left \lfloor E^{\rm H}/ e_{\rm q} \right \rfloor \right \} \right), && \text{if}\;\pi(s) = (I,H),\\
%&\mathbbm{1} \left( B' = {\rm min} \left \{b_{\rm max}, B - E^{\rm S} + \left \lfloor E^{\rm H}/ e_{\rm q} \right \rfloor \right \} \right), && \text{if}\;\pi(s) = (S,H),
%\end{aligned}
%\end{cases}
%\end{align}
%\begin{align}\label{eq:AoI_transprob}
%\P\left(A' \g A,\tau, \pi(s) \right) = \begin{cases}
%\begin{aligned}
%&\mathbbm{1}\left( A' = {\rm min}\left \{A_{\rm max}, \tau+1 \right \} \right),\; &&\text{if}\; \pi(s) = \left(a_1,T\right)\\
%&\mathbbm{1}\left( A' = {\rm min}\left \{A_{\rm max}, A+1 \right \} \right),\; &&\text{otherwise},
%\end{aligned}
%\end{cases}
%\end{align}
%\begin{align}\label{eq:tau_transprob}
%\P\left(\tau' \g \tau, \pi(s) \right) = \begin{cases}
%\begin{aligned}
%&\mathbbm{1}\left( \tau' = 1 \right),\; &&\text{if}\; \pi(s) = \left(S,a_2\right),\\
%&\mathbbm{1}\left( \tau' = {\rm min}\left \{\tau_{\rm max}, \tau+1 \right \} \right),\; &&\text{otherwise},
%\end{aligned}
%\end{cases}
%\end{align}
%where $\mathbbm{1}(\cdot)$ is the indicator function. 
The optimal policy $\pi^\star$ can be characterized using following Lemma.
\begin{lemma}
The policy $\pi^\star$ can be obtained by solving the following Bellman's equation for average cost MDPs \cite{bertsekas2011dynamic}
\begin{align}\label{belman_equation}
\bar{A}^\star + V(s) = \underset{a \in \nbbA(s)}{\rm min} Q(s,a), s \in \nbbS,
\end{align}
where $V(s)$ is the value function, $\bar{A}^\star$ is the achievable average AoI by $\pi^\star$ which is independent of the initial state $s(0)$, and $Q(s,a)$ is the expected cost due to taking action $a$ in state $s$, which is given by
\begin{align}\label{Q_func}
Q(s,a) = A + \sum\limits_{s' \in \nbbS} {\P(s' \g s, a) V(s')},
\end{align}
where $\P(s' \g s, a)$ can be computed using $(\ref{transprob})$. In addition, the optimal action taken at state $s$ can be evaluated as
\begin{align}\label{eq:optimal_policy}
\pi^\star(s) = {\rm arg} \underset{a \in \nbbA(s)}{\rm min} Q(s,a).
\end{align}
\end{lemma}

The value function $V(s)$ can be obtained iteratively using the VIA \cite{bertsekas2011dynamic}. Particularly, according to the VIA, the value function at iteration $k$, $k = 1, 2, \cdots$, is evaluated as 
\begin{align}\label{value_function_itern}
\nonumber V(s)^{(k)} & = \underset{a \in \nbbA(s)}{\rm min} Q(s,a)^{(k - 1)} \\
&= \underset{a \in \nbbA(s)}{\rm min} \left \{A + \sum\limits_{s' \in \nbbS} {\P(s' \g s, a) V(s')^{(k - 1)}}\right \},
\end{align}
where $s \in \nbbS$. Hence, $\pi^{\star}(s)$ at iteration $k$ is given by
\begin{align}\label{policy_itern}
\pi^{\star(k)}(s) = {\rm arg} \underset{a \in \nbbA(s)}{\rm min} Q(s,a)^{(k - 1)}.
\end{align}

Note that in each iteration of the VIA, the optimal action at each system state needs to be computed using (\ref{policy_itern}) (this is referred to as the policy improvement step). Under any initialization of value function $V(s)^{(0)}$, according to the VIA, the sequence $\left \{ V(s)^{(k)}\right\}$ converges to $V(s)$ which satisfies the Bellman's equation in (\ref{belman_equation}), i.e.,
%$\underset{k \rightarrow \infty}{\rm lim} V(s)^{(k)} = V(s)$.
%\vspace{-0.35 cm}
\begin{align}\label{conv}
\underset{k \rightarrow \infty}{\rm lim} V(s)^{(k)} = V(s).
\end{align}
%$\underset{k \rightarrow \infty}{\rm lim} V(s)^{(k)} = V(s)$.

In the next subsection, we will use the VIA to explore the structural properties of $\pi^\star$, which will be exploited to reduce the computational complexity of the VIA (as will be demonstrated in Remark \ref{rem:2}). Note that the obtained analytical results can be derived using Relative VIA (RVIA) as well \cite{bertsekas2011dynamic}.
\subsection{Structural Properties of the Optimal Policy}\label{sec: age_form}
\begin{lemma}\label{lem:1}
The value function $V(s)$ corresponding to $\pi^\star$ is: (i) non-decreasing w.r.t. $A$ and $\tau$, and (ii) non-increasing w.r.t. $B$, $g$ and $h$.
\end{lemma}
\begin{IEEEproof}
We first prove that $V(B,A,\tau,h,g)$ is non-decreasing w.r.t. $A$. Let $s\setminus x$ denote the combination of state $s$ variables excluding the variable $x$. Define $s_1 = (B_1,A_1,\tau_1,h_1,g_1)$ and $s_2 = (B_2,A_2,\tau_2,h_2,g_2)$ such that $A_1 \leq A_2$ and $s_1\setminus A_1 = s_2\setminus A_2$. Therefore, the goal is to show that $V(s_1) \leq V(s_2)$. Clearly, it is sufficient to show that the relation holds over all iterations of the VIA, i.e., $V(s_1)^{(k)} \leq V(s_2)^{(k)}, \forall k$. We prove that using mathematical induction as follows. For $k = 0$, the relation holds since we can choose the initial values $\{V(s)^{(0)}\}_{s \in \nbbS}$ arbitrary. Now, for an arbitrary value of $k$, we show that having $V(s_1)^{(k)} \leq V(s_2)^{(k)}$ leads to $V(s_1)^{(k + 1)} \leq V(s_2)^{(k + 1)}$. From (\ref{value_function_itern}) and (\ref{policy_itern}), $V(s_1)^{(k + 1)}$ and  $V(s_2)^{(k + 1)}$ are given, respectively, by
\begin{align}\label{v_s1_m+1}
\nonumber V(s_1)^{(k + 1)} & = A + \sum\limits_{s'_1 \in \nbbS}{\P(s'_1 \g s_1, \pi^{\star (k + 1)} (s_1)) V(s'_1)^{(k)}} \\
\nonumber & \overset{({\rm a})}{\leq} A + \sum\limits_{s'_1 \in \nbbS} {\P(s'_1 \g s_1, \pi^{\star (k + 1)} (s_2)) V(s'_1)^{(k )}} \\
& \overset{({\rm b})}{=} A + C \sum\limits_{g'_1} \sum\limits_{h'_1}{ V(\bar{B}_1,\bar{A}_1,\bar{\tau}_1,h'_1,g'_1)^{(k)}},
\end{align}
\begin{align}\label{v_s2_m+1}
\nonumber V(s_2)^{(k + 1)} &= A + \sum\limits_{s'_2 \in \nbbS} {\P(s'_2 \g s_2, \pi^{\star (k + 1)} (s_2)) V(s'_2)^{(k )}}\\
& \overset{({\rm b})}{=} A + C \sum\limits_{g'_2} \sum\limits_{h'_2}{ V(\bar{B}_2,\bar{A}_2,\bar{\tau}_2,h'_2,g'_2)^{(k)}},
\end{align} 
where step (a) follows since it is not optimal to take action $\pi^{\star(k + 1)}(s_2)$ in state $s_1$; step (b) follows from (\ref{eq:batt_evol})-(\ref{eq:tau_evol}) and (\ref{transprob}) where for a given $\pi^{\star(k + 1)}(s_2)$: 1) $\bar{B}_i$ and $\bar{\tau}_i$ are determined using (\ref{eq:batt_evol}) and (\ref{eq:tau_evol}), respectively, and 2) $\bar{A}_i$ is evaluated from (\ref{eq:AoI_evol}), $i \in \{1,2\}$. Note that $\bar{B}_1 = \bar{B}_2$ and $\bar{\tau}_1 = \bar{\tau}_2$ for $\pi^{\star(k + 1)}(s_2) \in \nbbA$ since we have $B_1 = B_2$ and $\tau_1 = \tau_2$. On the other hand, since $A_1 \leq A_2$, we can observe from (\ref{eq:AoI_evol}) that $\bar{A}_1 \leq \bar{A}_2$ for $\pi^{\star(k + 1)}(s_2) \in \nbbA$, and hence $V(\bar{B}_1,\bar{A}_1,\bar{\tau}_1,h'_1,g'_1)^{(k )} \leq V(\bar{B}_2,\bar{A}_2,\bar{\tau}_2,h'_2,g'_2)^{(k)}$. Therefore, $V(s_2)^{(k+1)}$ is greater than or equal to the expression in (\ref{v_s1_m+1}) which makes $V(s_1)^{(k + 1)} \leq V(s_2)^{(k + 1)}$ and indicates that the value function is non-decreasing w.r.t. $A$. Using the same approach, we can show that $V(B,A,\tau,h,g)$ is non-decreasing (non-increasing) w.r.t. $\tau$ ($B$). Finally, note that increasing $h$ ($g$) reduces $E^{\rm T}$ (increases $E^{\rm H}$), which increases the battery level at the next time slot and hence the value function is reduced. Therefore, $V(B,A,\tau,h,g)$ is non-increasing w.r.t. $h$ and $g$.

%Now, to prove that $V(B,A,\tau,g,h)$ is non-increasing w.r.t. $B$, let us assume $B_2 \leq B_1$ and $s_1\setminus B_1 = s\setminus B_2$. Hence, the objective is to show that $V(s_1) \leq V(s_2)$. Similar to the proof of part (i), we prove this relation using mathematical induction. In particular, we assume that $V(s_1)^{(m)} \leq V(s_2)^{(m)}$ holds for some $m$, and then show that it holds for $m + 1$ as well. Note that $V(s_1)^{(m + 1)}$ and  $V(s_2)^{(m + 1)}$ can be expressed as in (\ref{v_s1_m+1}) and (\ref{v_s2_m+1}), respectively. Since $A_1 = A_2$ and $\tau_1 = \tau_2$, we have $\bar{A}_1 = \bar{A}_2$ and $\bar{\tau}_1 = \bar{\tau}_2$ for $\pi^{\star(m + 1)}(s_2) \in \nbbA$. On the other hand, from (\ref{eq:batt_evol}), we have $\bar{B}_2 \leq \bar{B}_1$ for $\pi^{\star(m + 1)}(s_2) \in \nbbA$ since $B_2 \leq B_1$ by construction. Hence, $V(\bar{B}_1,\bar{A}_1,\bar{\tau}_1,g'_1,h'_1)^{(m )} \leq V(\bar{B}_2,\bar{A}_2,\bar{\tau}_2,g'_2,h'_2)^{(m )}$ which makes $V(s_1)^{(m + 1)} \leq V(s_2)^{(m + 1)}$ and indicates that the value function is non-increasing w.r.t. $B$. Finally, note that increasing $g$ ($h$) increases $E^{\rm H}$ (reduces $E^{\rm T}$) which leads to a larger amount of energy in the battery at the next time slot and hence a lower value function. This proves that $V(B,A,\tau,g,h)$ is non-increasing w.r.t. $g$ and $h$, and completes the proof of part (ii). 
\end{IEEEproof}
Using the monotonicity property of the value function, as demonstrated by Lemma \ref{lem:1}, the following Theorem characterizes some structural properties of the AoI-optimal policy $\pi^\star$.
\begin{theorem}\label{lem:2}
For any $s_1 = (B_1,A_1,\tau_1,h_1,g_1)$ and $s_2 = (B_2,A_2,\tau_2,h_2,g_2)$, the AoI-optimal policy $\pi^\star$ has the following structural properties: \\(i) When $B_1 \geq B_2$, $s_1\setminus B_1 = s_2\setminus B_2$ and $B_2 \geq b_{\rm max} - \left \lfloor E^{\rm H}_1 / e_{\rm q} \right \rfloor$, if $\pi^\star(s_1) = (I,H)$, then $\pi^\star(s_2) = (I,H)$.\\(ii) When  $B_1 \geq B_2$, $s_1\setminus B_1 = s_2\setminus B_2$ and $B_2 \geq b_{\rm max} - \left \lfloor E^{\rm H}_1 / e_{\rm q} \right \rfloor + E^{\rm S}$, if $\pi^\star(s_1) = (a_1,H)$, then $\pi^\star(s_2) = (a_1,H)$.\\(iii) When $A_2 \geq A_1$ and $s_1\setminus A_1 = s_2\setminus A_2$, if $\pi^\star(s_1) = (a_1,T)$, then $\pi^\star(s_2) = (a_1,T)$.\\(iv) When $\tau_2 \geq \tau_1$ and $s_1\setminus \tau_1 = s_2\setminus \tau_2$, if $\pi^\star(s_1) = (S,a_2)$, then $\pi^\star(s_2) = (S,a_2)$.
\end{theorem}
\begin{IEEEproof}
We first notice from (\ref{eq:optimal_policy}) that when $\pi^\star(s_1) = a$, we have $Q(s_1,a) - Q(s_1,a') \leq 0, \forall a'\in \nbbA(s_1)$. Hence, proving that $\pi^\star(s_1) = a$ leads to $\pi^\star(s_2) = a$ is equivalent to showing
\begin{align}\label{struc_prop}
Q(s_2,a) - Q(s_2,a') \leq Q(s_1,a) - Q(s_1,a'), \forall a' \neq a.
\end{align}

 For instance, to prove (i), we need to show that (\ref{struc_prop}) holds when $a = (I,H)$ and $a' \in\{(I,T),(S,H),(S,T)\}$. In the following, we prove part (i) while parts (ii), (iii) and (iv) can be proven similarly. According to (\ref{eq:batt_evol}), the next battery level for both states $s_1$ and $s_2$ when taking action $a = (I,H)$ is $b_{\rm max}$ since we have $B_1 \geq B_2$ and $B_2 \geq b_{\rm max} - \left \lfloor E^{\rm H}_1 / e_{\rm q} \right \rfloor$. Therefore, we have $Q(s_2,a) = Q(s_1,a)$ since $s_1\setminus B_1 = s_2\setminus B_2$, and showing that (\ref{struc_prop}) holds for (i) reduces to showing that $Q(s_1,a') \leq Q(s_2,a'), \forall a' \neq a$. Now, since $B_1 \geq B_2$, we note from (\ref{eq:batt_evol}) that the next battery level of $s_1$ is greater than or equal to the associated next battery level with $s_2$ for all possible values of $a' \neq a$. Therefore, based on Lemma \ref{lem:1} ($V(s)$ is non-increasing w.r.t. $B$), we have $Q(s_1,a') \leq Q(s_2,a'), \forall a' \neq a$ from (\ref{transprob}) and (\ref{Q_func}). This completes the proof of (i).
%  Particularly, from (\ref{transprob})-(\ref{eq:AoI_transprob}) and (\ref{Q_func}), we have
%
%\vspace{-0.35 cm}
%\small
%\begin{align}\label{Q_Si,T}
%Q(s_i,T) = A_i + C \sum\limits_{g'_i} \sum\limits_{h'_i}{ V(b_i - e^{\rm T}_i,1,g'_i,h'_i)},
%\end{align}
%\begin{align}\label{Q_Si,H}
%Q(s_i,H) = A_i + C \sum\limits_{g'_i} \sum\limits_{h'_i}{ V(b_{\rm max},{\rm min}\{A_{\rm max}, A_i + 1 \},g'_i,h'_i)},
%\end{align} 
%\normalsize
%where $i \in \{1,2\}$ and the next battery level in (\ref{Q_Si,H}) is equal to $b_{\rm max}$ since $b_1 + e^{\rm H}_1 \geq b_{\rm max}$ and $b_1 \leq b_2$. Since $s_1 \preceq s_2$ and based on Lemma 1, we have $V(b_1 - e^{\rm T}_{1},1,g'_1,h'_1) \geq V(b_2 - e^{\rm T}_{2},1,g'_2,h'_2)$ ($e^{\rm T}_1 \geq e^{\rm T}_2$) and $V(b_{\rm max},{\rm min}\{A_{\rm max}, A_2 + 1 \},g'_2,h'_2) \geq V(b_{\rm max},{\rm min}\{A_{\rm max}, A_1 + 1 \},g'_1,h'_1)$. Hence, (\ref{struc_prop}) holds for $a = T$ and $a' = H$, which completes the proof of (i).
\end{IEEEproof}
\begin{remark}\label{rem:1}
Theorem \ref{lem:2} demonstrates the threshold-based structure of the AoI-optimal policy $\pi^\star$ w.r.t. each of the system state variables. Specifically from (i) and (ii), we can see that $\pi^\star$ has a threshold-based structure w.r.t. $B$ when taking action $(I,H)$ for $B \geq b_{\rm max} - \left \lfloor E^{\rm H}_1 / e_{\rm q} \right \rfloor$ (when taking action $(a_1,H)$ for $B \geq b_{\rm max} - \left \lfloor E^{\rm H}_1 / e_{\rm q} \right \rfloor + E^{\rm S}$). For instance, for a fixed $s\setminus B$, if $B_{\rm th}$ is the maximum value of $B \geq b_{\rm max} - \left \lfloor E^{\rm H}_1 / e_{\rm q} \right \rfloor$ for which it is optimal to take an action $a = (I,H)$, then for all states $s$ such that $B \leq B_{\rm th}$, the optimal decision is $(I,H)$ as well. Similarly, from (iii) and (iv), we observe that $\pi^\star$ has a threshold-based structure w.r.t. $A$ and $\tau$ when taking actions $(a_1,T)$ and $(S,a_2)$, respectively. This essentially means that $\pi^\star$ aims to restrict the occurrence of the scenario of having a large AoI value at the destination node. In fact, in such a scenario, $\pi^\star$ would allocate a time slot for update packet transmission as soon as the source node has enough energy required for performing that action so that the average AoI at the destination node (expressed in (\ref{average_AoI})) is minimized.
\end{remark}
%In fact, in such scenario, $\pi^\star$ allocates the source node a time slot for an update packet transmission once it has enough energy required for performing that action such that the long-term average AoI at the destination node (expressed in (\ref{average_AoI})) is minimized.
One can also show that (\ref{struc_prop}) does not hold when $B_1 < B_2$ in parts (i) and (ii), $A_2 < A_1$ in part (iii) or $\tau_2 < \tau_1$ in part (iv). Because of this, it is not possible to discuss structural properties in this case.
\begin{remark}\label{rem:2}
Based on Remark \ref{rem:1}, the threshold-based structure of $\pi^{\star}$ w.r.t. the system state variables can be exploited to reduce the computational complexity of the VIA in terms of the number of required evaluations. More specifically, due to the threshold-based structure of $\pi^{\star}$, the optimal actions at some states can be directly determined based on the optimal actions taken at some other states without performing any evaluations. This, in turn, reduces the number of evaluations needed for the policy improvement step, and hence the computational complexity of the VIA is reduced. We refer the readers to \cite{hsu2019scheduling,zhou2018joint} for a detailed treatment of this point.
\end{remark}
%Recalling that the policy improvement step in each iteration of the VIA algorithm requires computing the optimal action at each system state using (\ref{policy_itern}), the computational complexity of characterizing $\pi^{\star}$ using the VIA is $\mathcal{O}\left(|\nbbS|\right)$, where $|\nbbS|$ is the cardinality of the system state space. Based on Remark \ref{rem:1}, the threshold-based structure of $\pi^{\star}$ w.r.t. the system state variables can be exploited to significantly reduce this complexity. More specifically, due to the threshold-based structure of $\pi^{\star}$, the optimal actions at some states can be directly determined based on the optimal actions taken at some other states without performing any evaluations. We refer the readers to \cite{hsu2019scheduling,zhou2018joint} for a detailed treatment of this point.
\begin{figure*}[t!]
\centerline{
\subfloat[]{\includegraphics[width=0.48\columnwidth]{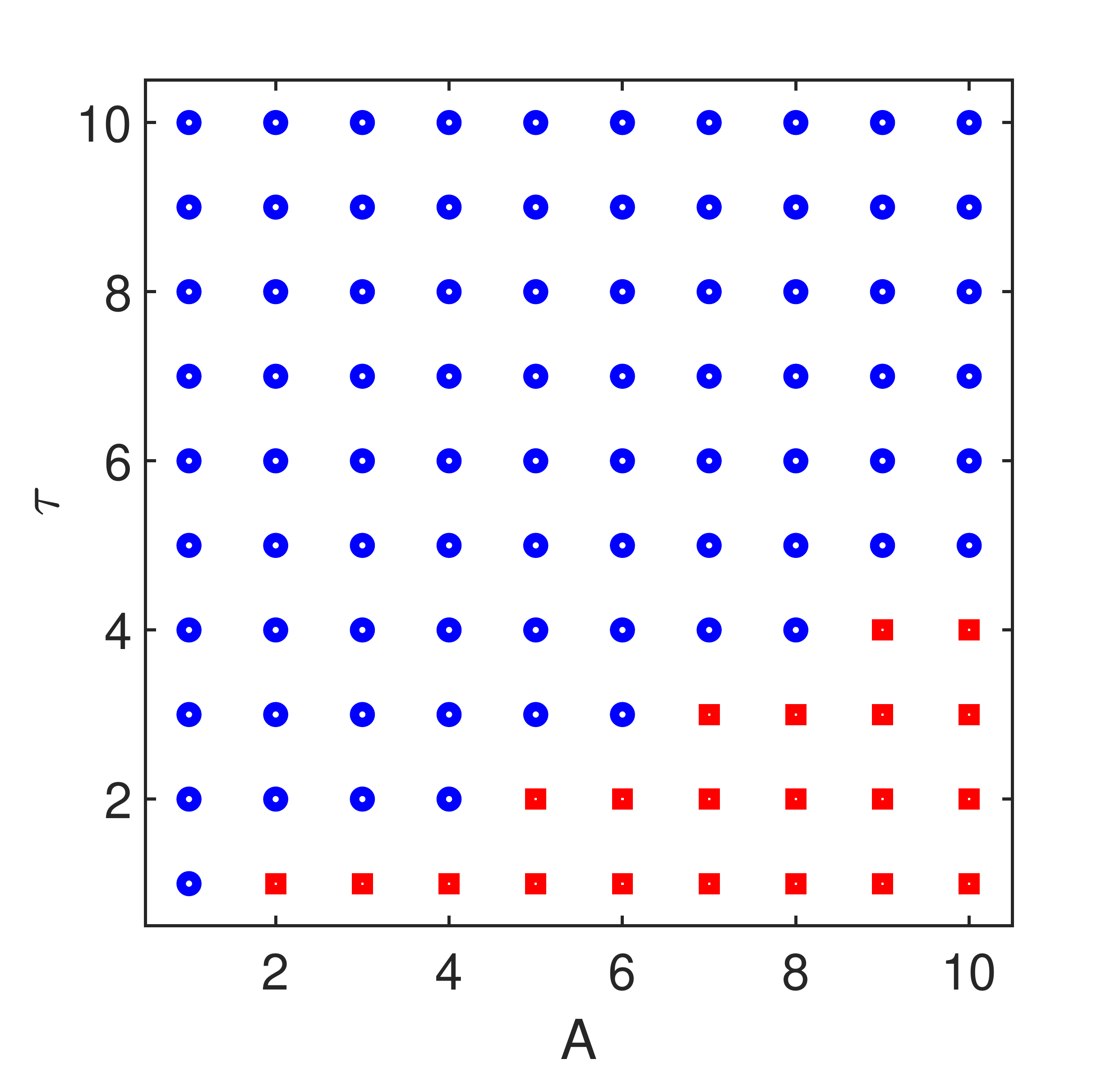}%
\label{f:a}} \hfil
\subfloat[]{\includegraphics[width=0.48\columnwidth]{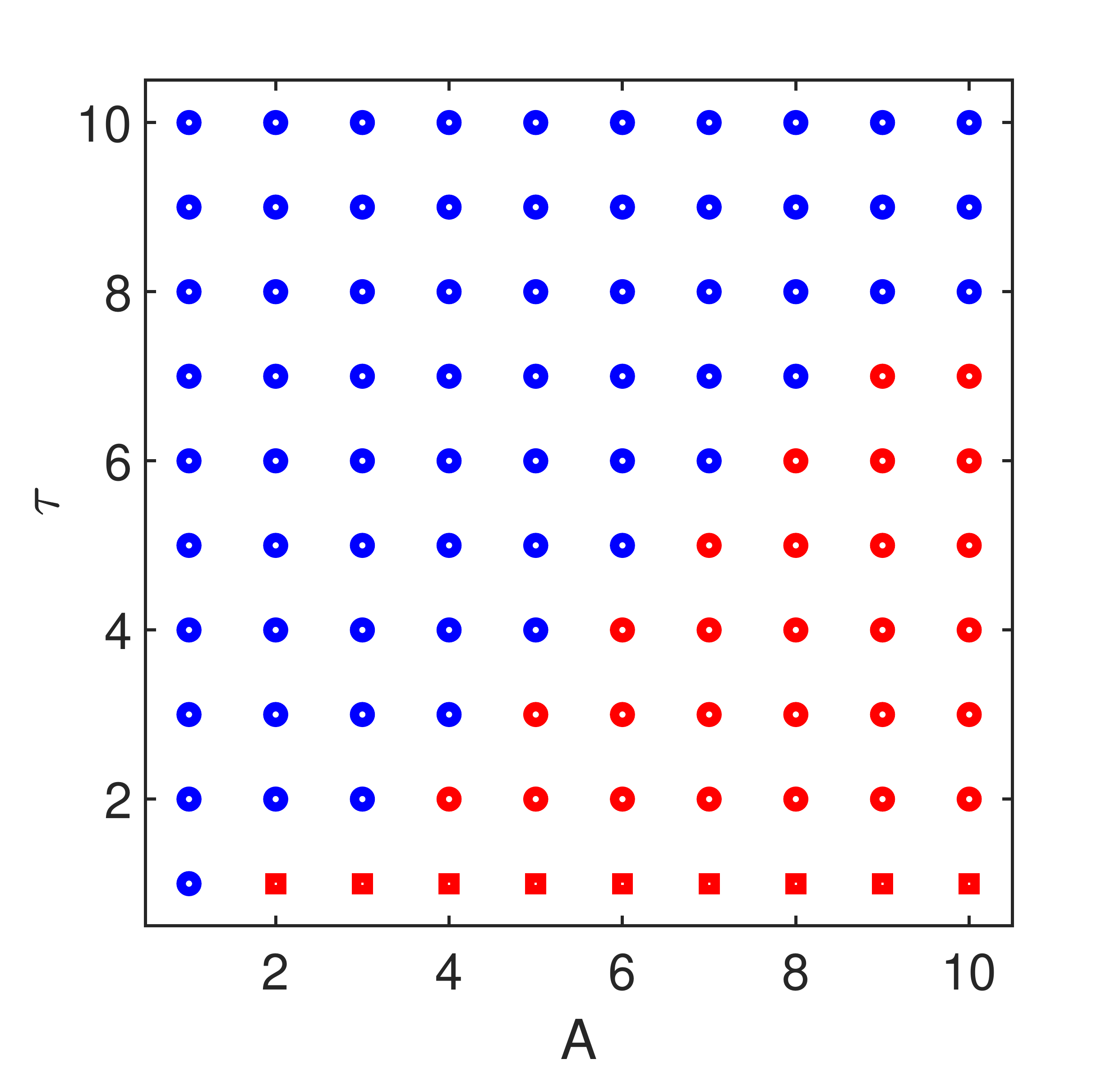}%
\label{f:b}} \hfil
\subfloat[]{\includegraphics[width=0.48\columnwidth]{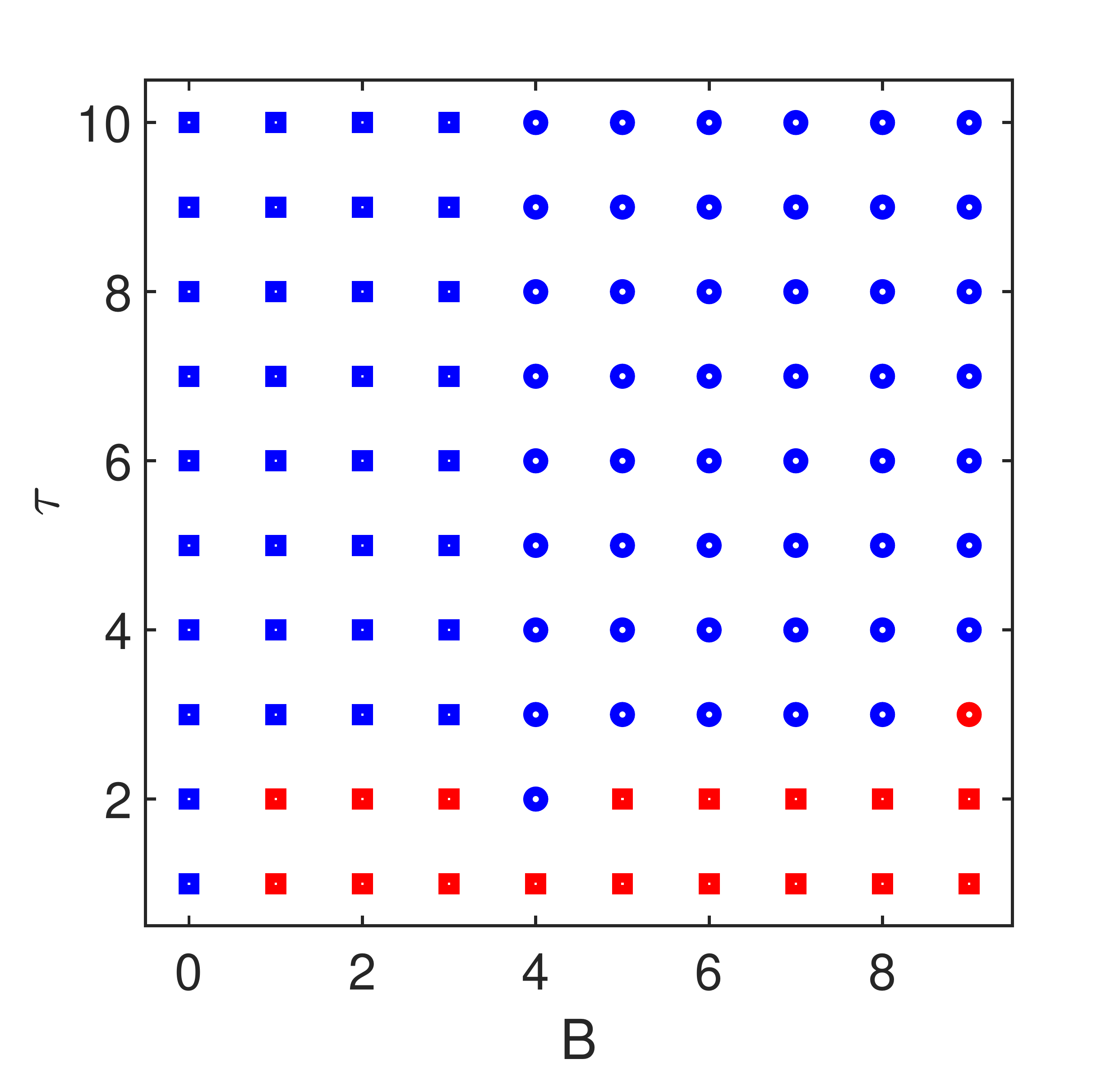}%
\label{f:c}}} \vfil
\centerline{
\subfloat[]{\includegraphics[width=0.48\columnwidth]{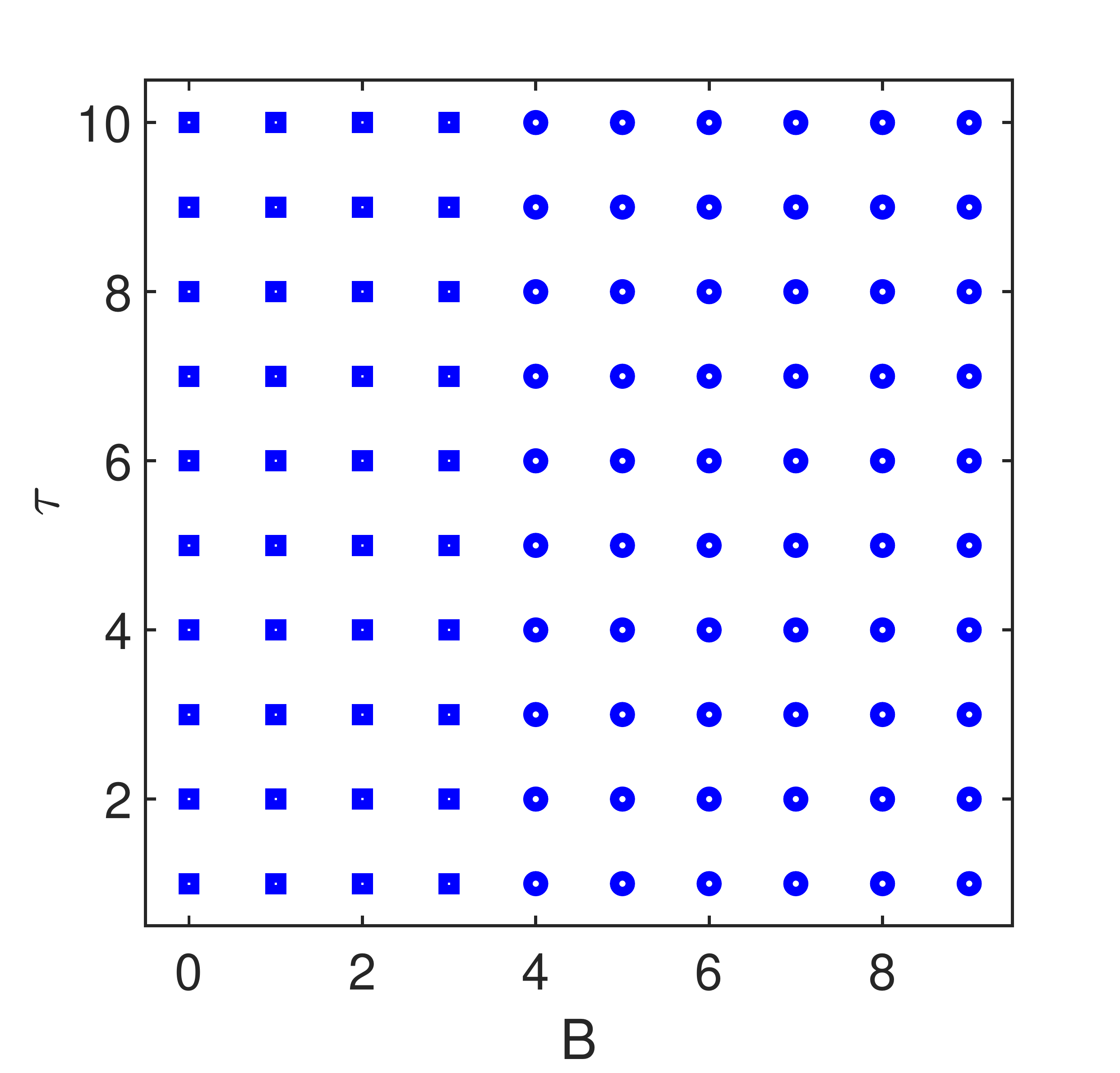}%
\label{f:d}} \hfil
\subfloat[]{\includegraphics[width=0.48\columnwidth]{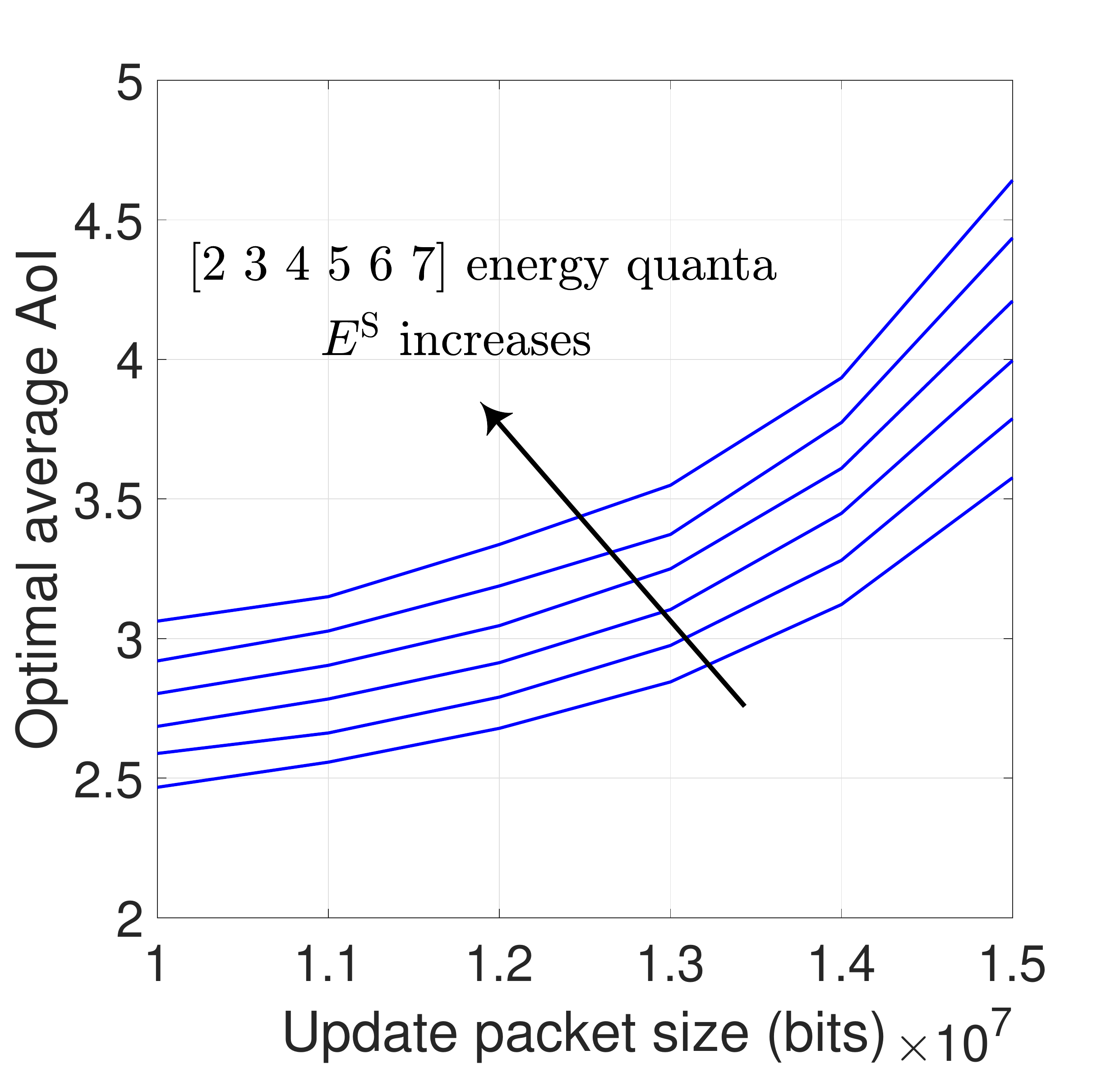}%
\label{f:e}} \hfil
\subfloat[]{\includegraphics[width=0.48\columnwidth]{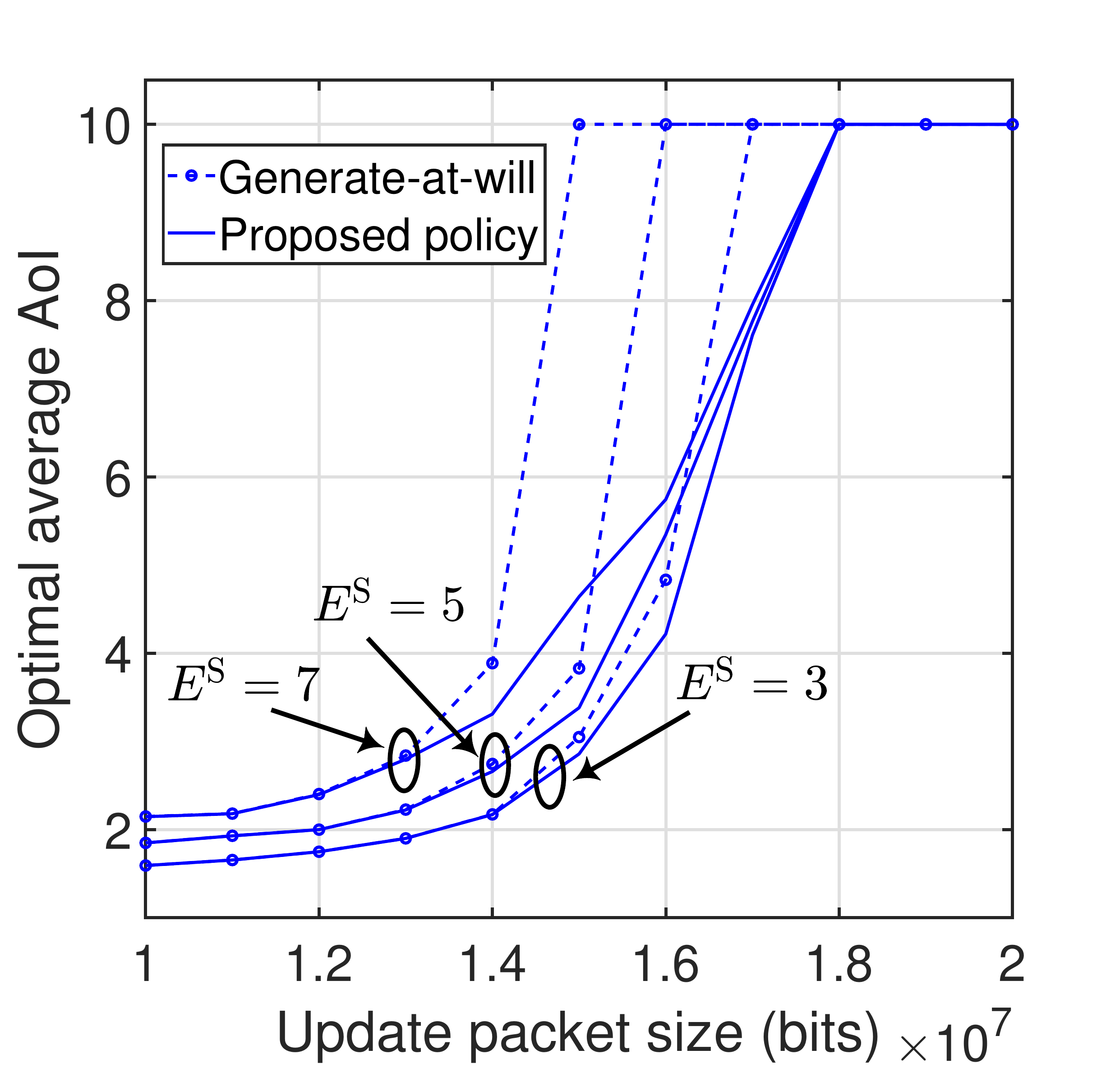}%
\label{f:f}}
} \caption{Structure of the AoI-optimal policy: (a) $E^{\rm S} = 3$ and $B = g = h = 5$, (b) $E^{\rm S} = 3$, $g = h = 5$, and $B = 9$, (c) $E^{\rm S} = 4$, $g = h = 6$, and $A = 5$, and (d) $E^{\rm S} = 4$, $g = h = 6$, and $A = 1$. System design insights: (e) Impact of $E^{\rm S}$ on the optimal achievable average AoI, and (f) Comparison between the performance of the proposed joint sampling and updating policy and that of the generate-at-will policy proposed in \cite{abd2019tcom}.}
\end{figure*} 
\section{Numerical Results}\label{sec:results}
%In this section, we first compare the structures of the AoI-optimal and throughput-optimal policies numerically and verify our analytical results derived in Sections \ref{sec: age_form} and \ref{sec: comparison}. Afterwards, we show the impact of system design parameters on the optimal achievable average AoI. 
We model the uplink and downlink channel power gains between the source and destination as $g = h = \delta \theta^2 d^{-\beta}$; $\delta$ is the gain of the signal power at a distance of $1$ meter, $d^{-\beta}$ models power law path-loss with exponent $\beta$, and $\theta^2\sim \exp(1)$ denotes the small-scale fading gain. Each state variable is discretized into $10$ levels. Considering a similar simulation setup to that of \cite{boshkovska2015practical}, we use $W = 1$ MHz, $d = 25$ meters, $P_{\rm t} = 37$ dBm, $P_{\rm max} = 12$ dBm, $\sigma^2 = - 95$ dBm, $M = 12$ Mbits, $B_{\rm max} = 0.3$ mjoules, $a = 1500$, $b = 0.0022$, $\delta = 4 \times 10^{-2}$, and $\beta = 2$. We also consider that the sensitivity of the power received at the RF energy harvesting circuit is $- 13$ dBm. Note that we use the red (blue) color to represent $a_2 = T$ $(a_2 = H)$ whereas the circle (square) marker to represent $a_1 = S$ $(a_1 = I)$.

First, from Figs. \ref{f:a}, \ref{f:b}, \ref{f:c} and \ref{f:d}, we can verify the analytical structural properties of $\pi^\star$ derived in Theorem \ref{lem:2}. For instance, we can observe from Figs. \ref{f:a} and \ref{f:b} that $\pi^{\star}$ has a threshold-based structure w.r.t. $A$ ($\tau$) when action $(a_1,T)$ (action $(S,a_2)$) is taken, as derived in parts (iii) and  (iv) of Theorem \ref{lem:2}. In addition, parts (i) and (ii) of Theorem \ref{lem:2} can be verified from Figure \ref{f:c}. For instance, since $\left \lfloor E^{\rm H}/ e_{\rm q}\right \rfloor = 9$ and $E^{\rm S} = 4$, we can see that: 1) since the optimal action at the point $(3,4)$ is $(I,H)$, it is optimal to take action $(I,H)$ at the points $(B,4), 0 \leq B \leq 3$ (part (i) in Theorem \ref{lem:2}), and 2) since the optimal action at the point $(9,4)$ is $(S,H)$, it is optimal to take action $(S,H)$ at the points $(B,4), 4 \leq B \leq 9$ (part (ii) in Theorem \ref{lem:2}). Second, the impact of $E^{\rm S}$ on $\pi^\star$ is revealed in Figs. \ref{f:a} and \ref{f:b}, where $\left \lceil E^{\rm T}/ e_{\rm q}\right \rceil = 2$. In particular, we discuss this impact in two different regimes: 1) the value of $E^{\rm S}$ is comparable with $B$ ($E^{\rm S}/B = 3/5$ in Fig. \ref{f:a}), and 2) $E^{S}$ is small w.r.t. $B$ ($E^{\rm S}/B = 3/9$ in Fig. \ref{f:b}). We observe that when $E^{s}$ is comparable with $B$ and $\tau$ is relatively large, it is optimal to take action $(S,H)$ and save energy that could be used for an update packet transmission for future packet transmissions when $\tau$ is small. Note that this insight can also be obtained for small values of $A$ (e.g., $A = 1$ in Fig. \ref{f:d}). 

Third, we show the impact of $M$ on the optimal achievable average AoI $(\bar{A}^\star)$ in Fig. \ref{f:e}. As expected, $\bar{A}^\star$ monotonically increases w.r.t. $M$ since the larger $M$, the larger is $E^{\rm T}$ required for its transmission. Finally, in Fig. \ref{f:f}, we demonstrate the importance of our proposed joint sampling and updating policy by comparing its achievable average AoI with that of the generate-at-will policy proposed in \cite{abd2019tcom}. The generate-at-will policy just decides whether to allocate each time slot for an update packet transmission or WET such that the update packets are only generated at the beginning of the time slots allocated for update packet transmissions. This means that the generate-at-will policy does not optimize the timing of update packet generations, and hence Fig. \ref{f:f} captures the impact of optimally generating update packets on $\bar{A}^\star$. We observe from Fig. \ref{f:f} that the achievable average AoI by our proposed policy significantly outperforms that of the generate-at-will policy \cite{abd2019tcom} especially when $M$ is large and/or when $E^{\rm S}$ is large. This happens since it becomes crucial in such cases to wisely decide the timing of update packet generations so that the energy available at the battery can be efficiently utilized to achieve a small value of average AoI.

\section{Conclusion}\label{sec:con}
%This letter studied the long-term average AoI minimization problem for wireless powered communication systems while taking into account the costs of generating status updates at the source nodes. The problem was modeled as an average cost MDP for which its corresponding value function was shown to be monotonic w.r.t. state variables. We analytically demonstrated the threshold-based structure of the AoI-optimal policy w.r.t. state variables. The numerical results verified the analytical findings as well as showed the impact of the energy required for generating a status update at the source (w.r.t. its battery level) on the structure of the AoI-optimal policy.
This paper has studied the long-term average AoI minimization problem for wireless powered communication systems while taking into account the costs of generating status updates at the source nodes. The problem was modeled as an average cost MDP for which its corresponding value function was shown to be monotonic w.r.t. state variables. We analytically demonstrated the threshold-based structure of the AoI-optimal policy w.r.t. state variables. Our numerical results revealed that when the energy required for an update packet generation is comparable with the energy available in the battery, the optimal action mainly depends on the time elapsed since the generation of the current packet available at the source. In particular, it is optimal to generate a new update packet if the current packet available at the source was generated from a relatively long time ago.  
%the optimal action mainly depends on the time passed since the generation of the current packet available at the source. In particular, it is optimal to generate a new update packet if the current packet available at the source was generated from a relatively long time ago. 
%Furthermore, our results demonstrated that the optimal achievable average AoI is monotonically increasing w.r.t. the size of update packets and/or the amount of energy required for an update packet generation. 
Our results also demonstrated the importance of optimally generating status updates by showing that the performance of our proposed joint sampling and updating policy significantly outperforms that of the generate-at-will policy in terms of the achievable average AoI. A promising avenue of future work is to extend our analysis and results to the scenario with multiple source nodes. Given the prohibitive complexity of the problem resulting from the extreme curse of dimensionality in the state space of its associated MDP, it is difficult to tackle it with conventional approaches. A feasible option it to use deep reinforcement learning-based algorithms to reduce the complexity of the state space while learning the optimal policy at the same time.

\bibliographystyle{IEEEtran}
\bibliography{Manuscript_v01}
\end{document}